\documentclass[aps,pre,amsmath,amssym,amsthm,superscriptaddress,reprint, nofootinbib,longbibliography]{revtex4-2}
\usepackage[colorlinks=true,urlcolor=blue,citecolor=blue,linkcolor=blue]{hyperref}

\usepackage{graphicx}
\usepackage{float}
\usepackage[]{placeins} 
\usepackage{dcolumn}
\usepackage{kotex}
\usepackage{natbib}
\usepackage{color}
\usepackage{amsfonts}
\usepackage{hyperref}
\usepackage{algorithm}
\usepackage{algcompatible}
\usepackage{newfloat}
\usepackage{makecell}
\usepackage{bm}
\usepackage{orcidlink}

\usepackage{setspace}
\usepackage{multirow}
\usepackage{sidecap}
\usepackage{enumerate}

\definecolor{mygray}{gray}{0.5}

\usepackage{xcolor}
\definecolor{tomato}{HTML}{FF6347}          
\definecolor{dodgerblue}{HTML}{1E90FF}      

\begin{document}

\title{Role of volatility mixing in wealth condensation transition}

\author{Jaeseok Hur\orcidlink{0009-0007-1976-275X}}
\affiliation{Department of Physics, Korea Advanced Institute of Science and Technology, Daejeon 34141, Korea}

\author{Meesoon Ha\orcidlink{0000-0003-1279-5325}}
\email[Contact author: ]{msha@chosun.ac.kr}
\affiliation{Department of Physics Education, Chosun University, Gwangju, 61452, Korea}

\author{Hawoong Jeong\orcidlink{0000-0002-2491-8620}}
\email[Contact author: ]{hjeong@kaist.edu}
\affiliation{Department of Physics, Korea Advanced Institute of Science and Technology, Daejeon 34141, Korea}
\affiliation{Center of Complex Systems, KAIST, Daejeon 34141, Korea}

\date{\today}

\begin{abstract}
We study the role of heterogeneous volatility in a networked wealth dynamics model and its impact on the wealth condensation transition. Extending the Bouchaud–M{\'e}zard framework, we introduce binary volatility in networks and investigate how its configuration affects the effective power-law tail exponent of the wealth distribution. Using a stochastic block model, we control the mixing between volatility groups and show that the effective exponent is governed not only by the global parameter $\Lambda=2J/\beta^2$ but also by the volatility configuration in the network. We find that local interactions between nodes with different volatility induce the neutralization of group-wise exponents, which lowers the aggregate tail exponent and yields a condensation transition across $\gamma_{\rm c}=2$. Our results identify volatility mixing as another control mechanism for wealth condensation and highlight the importance of noise heterogeneity in nonequilibrium systems on networks.
\end{abstract}

\maketitle

\section{Introduction}
\label{sec:intro}

Heavy-tailed wealth distributions are the hallmark of socioeconomic inequality, where the tail exponent $\gamma$ plays a central role in determining the degree of concentration. In the context of wealth dynamics, Bouchaud and M{\'e}zard~\cite{bouchaud2000wealth} (BM) established a paradigmatic framework, in which the stationary wealth distribution exhibits a power-law tail exponent $\gamma$ that depends only on the ratio of interaction strength to squared volatility, $\Lambda \equiv 2J/\beta^2$ (see Fig.~\ref{fig1}), both in the mean-field (MF) case and Erd\H{o}s--R{\'e}nyi (ER) networks~\cite{Erdos:1959:pmd}. This framework suggests that the condensation transition of wealth dynamics, signaled by $\gamma\le 2$, is governed by a single control parameter.

However, real systems are rarely homogeneous: individuals may differ not only in growth rates but also in volatility. Although heterogeneity in growth has been studied~\cite{bernard2026mean,hur2026anomaly}, the effect of heterogeneous volatility remains largely unexplored. This is also important because volatility represents intrinsic fluctuations in wealth and can interact with the network structure in ways that cannot be reduced to a single effective parameter.

In this paper, we investigate the effect of volatility mixing on the tail exponent of wealth distributions and the condensation transition in the BM model. In the context of a stochastic block model (SBM), we tune the mixing between high- and low-volatility groups while keeping degree statistics nearly unchanged. This allows us to isolate the volatility configuration effect itself.
We show that the effective tail exponent ($\hat{\gamma}$) is determined not only by the ratio $\Lambda$ but also by the allocation of volatility in the network. 
In particular, mixing between nodes with different volatility induces the neutralization of group-wise effective tail exponents, lowers the aggregate effective tail exponent, and eventually drives the system to the threshold $\gamma_{\rm c}=2$ for a condensation transition (see Fig.~\ref{fig1}). These results identify volatility configuration as an additional control mechanism for condensation in nonequilibrium networked systems. 

The remainder of this paper is organized as follows: In Sec.~\ref{sec:model}, we introduce the heterogeneous BM model and analyze its properties in the homogeneous limit ($\Delta\beta \to 0$) for both the MF case and ER networks. Using the SBM, we relate the volatility-volatility assortativity $\mathcal{A}$ to the probability of interblock connections $q$. In Sec.~\ref{sec:results}, we clarify the dependence of $\hat{\gamma}$ on $q$ and establish a configuration-induced wealth condensation transition. Finally, in Sec.~\ref{sec:conclusion}, we summarize our findings with some remarks on broader implications and possible directions for future study. More detailed explanations and tests are also provided in Appendixes \ref{detail} and \ref{complementary}.

\section{Model}
\label{sec:model}

The power-law-tailed distribution of normalized wealth $c$ is characterized by
\begin{align}
\rho(c)\sim c^{-\gamma} \quad \text{for large } c,
\label{eq1}
\end{align}
where $\gamma$ is a power-law tail exponent. In particular, for $\gamma \leq 2$, the mean of an ideal tail sector diverges, and a condensation occurs. The threshold $\gamma_{\rm c} = 2$ marks the condensation transition point, at which the inverse participation ratio (IPR), an order parameter~\cite{bouchaud1997universality}, and the Gini index become singular.

We introduce binary volatility into the BM model. A stochastic differential equation (SDE) for the heterogeneous BM (HBM) model is given by
\begin{align}
    dc_i=\frac{J}{\langle{k}\rangle}\sum_{j=1}^{N}a_{ij}(c_j-c_i)dt+\beta_i c_idW_{t,i},
    \label{eq2}
\end{align}
where $c_i$ denotes the normalized wealth~\cite{SDE-normalizedwealth} of node $i$, $t$ is time, $W_{t,i}$ is a Wiener process associated with node $i$ at time $t$, $\beta_i=\beta_{\pm}=\beta\pm\Delta\beta$ is the volatility of node $i$, $J$ is the interaction strength between connected nodes, $\mathbf{a}$ is the adjacency matrix of a given network with elements $a_{ij}$, $\langle{k}\rangle$ is the mean degree, and $N$ is the total number of nodes. Throughout this paper, Eq.~\eqref{eq2} is interpreted in the It{\^o} sense. In contrast to the homogeneous BM model, the volatility configuration $\bm{\beta}=(\beta_1,\dots,\beta_N)$ for a given network plays the role of a quenched disorder and also influences the wealth distribution.

\begin{figure}[t]
\includegraphics[width=\columnwidth]{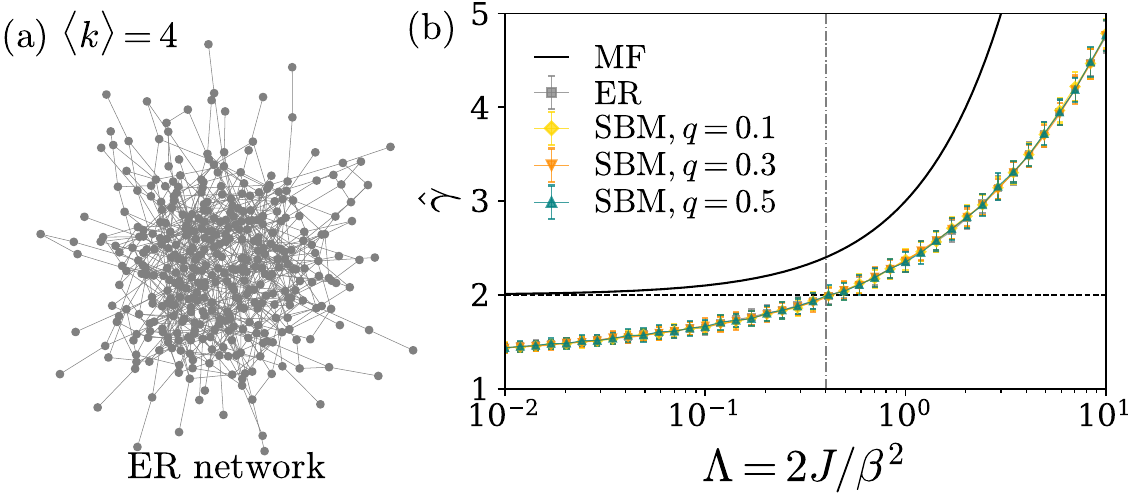}
    \caption{(a) Visualization of the ER network with mean degree $\langle k\rangle=4$. (b) Effective tail exponent $\hat{\gamma}$ of wealth distributions in the BM model for various network types as a function of $\Lambda=2J/\beta^2$. The solid line shows the analytical baseline for the MF case. The other cases show numerical simulation results, where symbols and error bars indicate means and standard deviations, respectively, over 128 runs. The mean degree is $\langle{k}\rangle=4.00\pm0.027$. For all cases, $J=10^{-3}$ and $N=10^4$.}
    \label{fig1}
\end{figure}

For a complete network ($a_{ij}=1-\delta_{ij}$), Eq.~\eqref{eq2} reduces, 
in the thermodynamic limit ($N\to\infty$), to 
\begin{align}
dc_i = J(1-c_i)dt + \beta_i c_idW_{t,i}.
\label{eq-complete}
\end{align}
In the homogeneous limit of $\Delta\beta\to0$, Eq.~\eqref{eq-complete} becomes the MF case of the BM model. The stationary distribution of the MF-BM model is an inverse-gamma distribution with tail exponent $\gamma = 2 + 2J/\beta^2$. Even when $\Delta\beta\neq0$, each volatility group still follows the corresponding MF result of the BM model because Eq.~\eqref{eq-complete} depends only on $c_i$ and $\beta_i$. The aggregate wealth distribution is therefore a mixture of two inverse-gamma  distributions with exponents $\gamma_{\pm}=2+2J/\beta_\pm^2$. Hence, in the MF limit, the group-wise tail exponents remain unchanged despite the presence of heterogeneous volatility.

For ER networks, the power-law tail of $\rho(c)$ has also been numerically confirmed. Moreover, the effective tail exponent $\hat{\gamma}$ does not depend on the individual values of $J$ and $\beta^{2}$, but only on their ratio~\cite{bouchaud2000wealth}. For convenience, we denote it as the control parameter $\Lambda$:
\begin{align}
\Lambda \equiv 2J/\beta^2.
\end{align}

Figure~\ref{fig1} shows the dependence of $\hat{\gamma}$ on $\Lambda$ for various cases. Although the MF case of the BM model always yields $\gamma>2$ and therefore no condensation, the BM model on ER networks can exhibit $\hat{\gamma}\leq2$ and a wealth condensation transition. As shown in Fig.~\ref{fig1}(b), for $\langle{k}\rangle = 4$, the transition occurs around $\Lambda\approx 0.4$. It is noted that the effective tail exponent $\hat{\gamma}$ is estimated using a maximum-likelihood procedure together with a data-driven tail cutoff (Appendix~\ref{detail}), and finite-size effects in tail fitting are assessed separately (Appendix~\ref{complementary}).

However, for the case of binary volatility ($\Delta\beta\neq 0$) with $\beta_{\pm}=\beta\pm\Delta \beta$, corresponding group-wise exponents $\hat{\gamma}_{\pm}$ differ substantially from the values obtained simply by substituting $\beta_\pm$ into the result for the homogeneous limit. This difference arises from local interactions between nodes with heterogeneous volatility, which distinguishes the networked model from the MF case.
Since the values of $\hat{\gamma}_{\pm}$ depend on how $\beta_i$ is allocated to the nodes of a given network, we study this configuration effect on wealth condensation in the HBM model.

To characterize the configurational property relevant to the effective tail exponent, we consider the volatility-volatility assortativity $\mathcal{A}$ of heterogeneous volatility, which measures the tendency of nodes with similar volatility to be connected for a given network:
\begin{align}
\mathcal{A}\equiv\frac{\mathrm{Cov}(\beta,\beta')}{\sqrt{\mathrm{Var}(\beta)\mathrm{Var}(\beta')}},
\label{eq3}
\end{align}
where $\beta$ and $\beta'$ denote the volatility of two connected nodes, respectively. In principle, $-1\leq\mathcal{A}\leq1$, but the attainable bound depends on the details of the network structure~\cite{cinelli2020network,degree-degree}. When $N_+=N_-$, the random permutation of binary volatility values for an ER network yields $\mathcal{A}\approx0$.

Although one may adjust $\mathcal{A}$ repeatedly by swapping the volatility values of a pair of nodes from an arbitrary initial configuration, this approach cannot go beyond the structural limit imposed by the given network. As a result, it becomes difficult to independently control other configurational properties, such as the degree--volatility correlation~\eqref{eq-C1} (see Appendix~\ref{Appendix-C3} for details). These limitations can be naturally overcome by introducing a stochastic block model (SBM).

\begin{figure}[t]
\includegraphics[width=0.9\columnwidth]{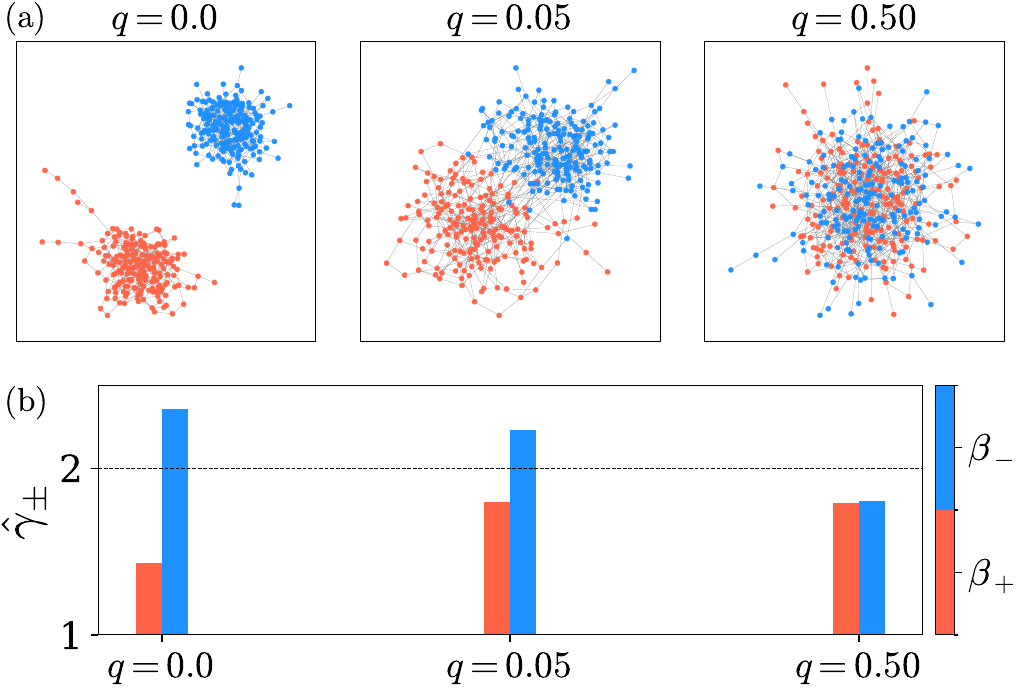}
\caption{(a) Three illustrations of SBM used for HBM model simulations. Here, $q$ is the probability of interblock connections, and node colors represent higher ($\color{tomato}\bullet$, $\beta_+$) and lower ($\color{dodgerblue}\bullet$, $\beta_-$) volatility. (b) Corresponding group-wise effective tail exponents $\hat{\gamma}_{\pm}$ for $q=\{0.0,\ 0.05,\ 0.50\}$ at $(\Lambda_1,\Lambda_2)=(0.01,1)$.}
\label{fig2}
\end{figure}

In this paper, we consider an SBM consisting of two equally sized blocks, $a$ and $b$. Its block connection matrix is given by
\begin{align}
    P_{ab}=
    \begin{bmatrix}
        p(1-q) & pq \\
        pq & p(1-q)
    \end{bmatrix},
    \label{eq4}
\end{align}
where $p$ denotes the overall probability of connection between nodes in the SBM, and $q$ controls the probability of connections between blocks [see Fig.~\ref{fig2}(a)]. 

For a given mean degree $\langle{k}\rangle$, the corresponding ER network has the connection probability $p_0=\langle{k}\rangle/(N-1)$, while the SBM has $p\approx2p_0$ in the large-$N$ limit and the mean degree $\langle k\rangle$ is independent of $q$. In particular, when $q=1/2$, the SBM is equivalent to an ER network in which every pair of nodes is connected with probability $p/2$. The assignment of $\beta_+$ and $\beta_-$ to two blocks yields
\begin{align}
\mathcal{A}=1-2q.
\label{eq5}
\end{align}
Hence, the probability of interblock connection $q$ serves as a convenient control parameter for adjusting $\mathcal{A}$ without significantly altering the overall network structure.

%
\begin{figure*}[]
\includegraphics[width=\textwidth]{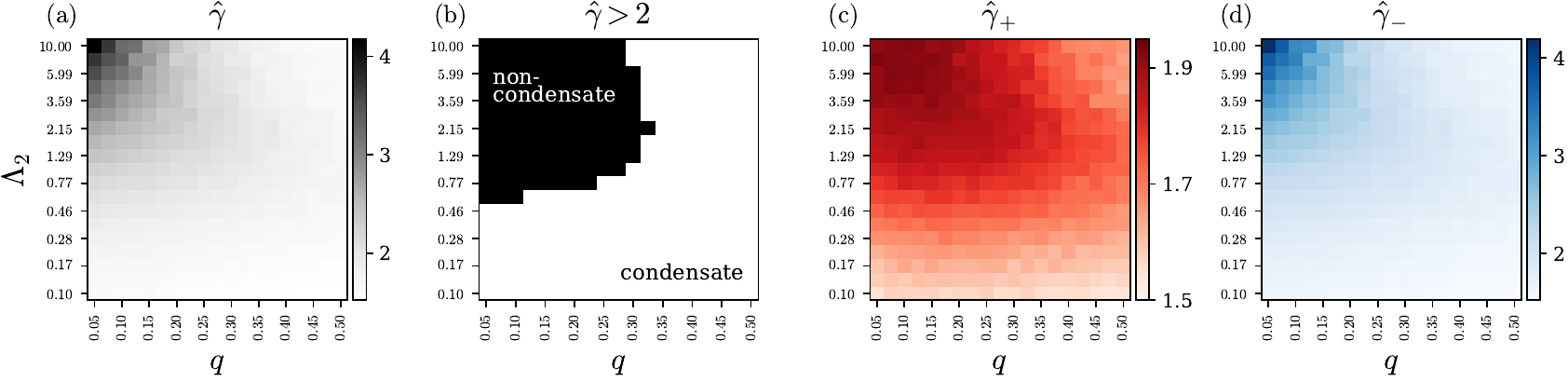}
\caption{(a) Aggregate effective tail exponent $\hat{\gamma}$ in the plane of $(q,\Lambda_2)$ at fixed $\Lambda_1=10^{-2}$. (b) Phase diagram of condensed ($\hat{\gamma}\leq2$) and noncondensed ($\hat{\gamma}>2$) regimes in the plane of $(q,\Lambda_2)$. (c) and (d) Group-wise effective tail exponents $\hat{\gamma}_{\pm}$ in the plane of $(q,\Lambda_2)$, respectively. In all panels, the underlying network is an SBM with $\langle{k}\rangle=4.00\pm 0.027$, and numerical simulations are performed for $\Lambda_1=10^{-2}$, $J=10^{-3}$, and $N\approx 10^4$. Here, estimated exponents are averaged over 128 independent runs.}
\label{fig3}
\end{figure*}

\section{Results}
\label{sec:results}

In the SBM, a wealth distribution exhibits nearly the same effective tail exponent $\hat{\gamma}$ as that in the ER network in the homogeneous limit of $\Delta\beta\to0$ since the degree distributions of the two networks are nearly identical. However, for $\Delta\beta\neq 0$, $\hat{\gamma}$ strongly depends on $q$. 

At $q=0$, two blocks form completely separated ER networks. Consequently, each group has an effective tail exponent $\hat{\gamma}_{\pm}$ corresponding to $\Lambda_{1,2}=2J/\beta_{\pm}^{2}$. However, for $q>0$, local interactions between nodes with different volatility modify $\hat{\gamma}_{\pm}$. Specifically, the gap between two exponents decreases, producing a neutralization effect (see Fig.~\ref{fig2}).
This trend is consistent with the form of the interaction term in Eq.~\eqref{eq2}, which redistributes wealth in proportion to the difference in normalized wealth between connected nodes. 

As $q$ increases, two nodes with different volatility interact more frequently, and the difference between two group-wise effective tail exponents decreases. Since this neutralization does not occur in the MF case, it originates from local interactions on the network.

In the homogeneous BM model, $\hat{\gamma}$ is determined solely by the control parameter $\Lambda$, while in the heterogeneous BM model, by contrast, the configurational property of volatility acts as an additional control dimension that determines the aggregate effective exponent $\hat{\gamma}$.
In order to demonstrate this, we fix $\Lambda_1 = 2J/\beta_+^2$ and construct a phase diagram in the plane of $(q,\Lambda_2)$, where $\Lambda_2 = 2J/\beta_-^2$. The value of $\hat{\gamma}$ extracted from the whole distribution $\rho(c)$ varies not only with $\Lambda_2$ but also with $q$. Moreover, for the appropriate values of $(q,\Lambda_1,\Lambda_2)$, the system undergoes a wealth condensation transition at $\gamma_{\rm c}=2$, indicating a phase boundary in the plane of $(q,\Lambda_2)$. Figure~\ref{fig3} shows that the larger $q$ is, the smaller $\hat{\gamma}$ becomes, and the overall inequality of the system becomes enhanced.

The neutralization of $\hat{\gamma}_{\pm}$ does not affect the total distribution symmetrically. In all parameter regimes that we tested, group-wise distributions $\rho_{\pm}(c)$ show that the tail population is larger in the low-volatility group than in the high-volatility group. Consequently, the tail of the aggregate distribution $\rho(c)$ is governed primarily by the tail behavior of the $\beta_-$ group (see Fig.~\ref{fig3}). This is also reflected in the phase diagrams. Because the low-volatility group contributes much more strongly to the tail, the wealth shares of the two groups also differ substantially (see Fig.~\ref{fig-A3}). This gap in wealth shares decreases as $q$ increases, indicating that interblock connections hinder the net transfer of wealth from the high-volatility group to the low-volatility group.

Lastly, we explore the HBM model beyond binary volatility and symmetric two-block SBM settings, in terms of the following three cases.
(i) {\it Continuous volatility within each block}. Instead of assigning a single volatility value to each block, we draw the volatility from a uniform distribution within each block:
\begin{align}
    \beta_a &\sim \mathrm{Uniform}\left(\beta_+ - \delta\beta,\beta_+ + \delta\beta\right), \nonumber \\
    \beta_b &\sim \mathrm{Uniform}\left(\beta_- - \delta\beta,\beta_- + \delta\beta\right).
\end{align}
Thus, the mean volatility in the two blocks is still given by $\beta_+$ and $\beta_-$, respectively. This setup reduces to the previous one of binary volatility in the limit $\delta\beta\to 0$. We set $\delta\beta = |\beta_+ - \beta_-|/2$ and examine the resulting group-wise wealth distributions. 
(ii) {\it Asymmetric block sizes}. We consider the case of different block sizes, $N_a\neq N_b$. To control the mean degree $\langle{k}\rangle$ in the limit of $N\to\infty$, we use the following block matrix:
\begin{align}
    P_{ab}=
    \begin{bmatrix}
    p_a(1-q) & \left(\frac{p_a+p_b}{2}\right)q \\
    \left(\frac{p_a+p_b}{2}\right)q & p_b(1-q)
    \end{bmatrix},
\end{align}
where $p_a=\langle{k}\rangle/(N_a-1)$ and $p_b=\langle{k}\rangle/(N_b-1)$. 
(iii) {\it Symmetric multiblock SBM}.  Finally, we consider a symmetric SBM with $M$ blocks of equal size. To control the mean degree $\langle{k}\rangle$ in the limit $N\to\infty$, we use the following block matrix:
\begin{align}
    P=
    \begin{bmatrix}
    p(1-q) & \frac{pq}{M-1} & \cdots & \frac{pq}{M-1} \\
    \frac{pq}{M-1} & p(1-q) & \cdots & \frac{pq}{M-1} \\
    \vdots & \vdots & \ddots & \vdots \\
    \frac{pq}{M-1} & \frac{pq}{M-1} & \cdots & p(1-q)
    \end{bmatrix},
\end{align}
where $p=\langle{k}\rangle/(N/M-1)$. The case $M=2$ reduces to the symmetric two-block setup considered mainly in our study. In numerical simulations, we examine a symmetric four-block SBM and assign volatility values to the four blocks that are equally spaced on a logarithmic scale.

The only difference is that the assortativity $\mathcal{A}$ is no longer represented by Eq.~\eqref{eq5}, although it remains a decreasing function of $q$. Moreover, the degree--volatility correlation remains nearly zero, except for the asymmetric case, leaving assortativity as the dominant influence.

\begin{figure}[b]
\includegraphics[width=\columnwidth]{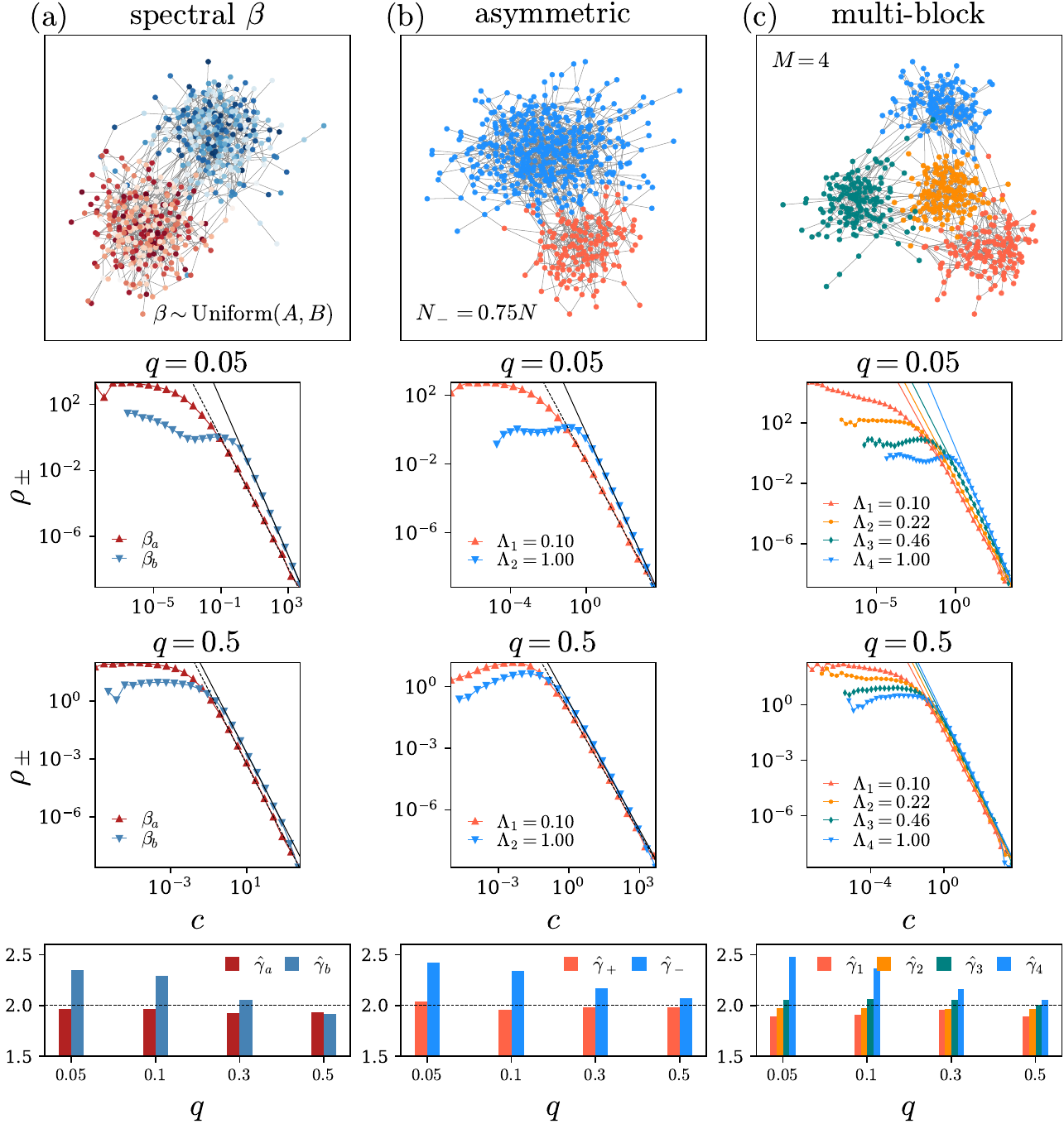}
\caption{From top to bottom, the panels present schematic illustrations for modified experimental setups (top), group-wise normalized wealth distributions $\rho_{\pm}(c)$ for $q=\{0.05,0.5\}$ (middle), and group-wise effective tail exponents $\hat{\gamma}_{\pm}$ as functions of $q$ (bottom), respectively. From left to right, the columns show the corresponding numerical results for (a) continuous volatility distributions within each block $(\beta_a,\beta_b)$, (b) an SBM with asymmetric block sizes ($N_a \neq N_b$), and (c) a symmetric multiblock SBM with $M=4$, respectively.}
\label{fig4}
\end{figure}

The numerical results for these three cases are shown in Fig.~\ref{fig4}. We emphasize that these results also exhibit the neutralization of the tail exponents and the dominance of the low-volatility group in the tail population, demonstrating that our findings for the effect of volatility mixing on wealth condensation transitions are not limited to binary volatility and symmetric two-block settings.

\section{Conclusion}
\label{sec:conclusion}

We investigated the role of heterogeneous volatility in the Bouchaud--M{\'e}zard model and showed that the configuration of volatility fundamentally alters the statistics of wealth distributions. In contrast to the homogeneous case, in which the tail exponent is uniquely determined by the control parameter $\Lambda$, we found that the effective exponent $\hat{\gamma}$ depends on the volatility mixing structure, controlled here by the probability of interblock connection $q$. This configurational effect is absent in the MF limit and emerges only when nodes with heterogeneous volatility interact locally on a sparse network.

A key mechanism underlying this behavior is the neutralization of group-wise tail exponents through local exchange between nodes with different volatility. As volatility mixing becomes more frequent, the disparity between subpopulations becomes reduced at the group level, yet the aggregate tail becomes heavier because the overall effective exponent decreases. In this way, wealth condensation can be induced not only by tuning a global parameter, such as $\Lambda$, but also by changing the configurational correlations in the noise intensity. The phase diagram in the plane of $(q,\Lambda_2)$ and the complementary robustness checks in Appendix~\ref{complementary} support this interpretation.

Our findings suggest a broader perspective on nonequilibrium systems with heterogeneous noise. Although earlier examples, such as active matter~\cite{vicsek1995novel,bonilla2019active} and the random-field Ising model~\cite{imry1975random,ahrens2011critical}, emphasize correlations in the noise itself, our results show that correlations in noise amplitude (i.e., correlated volatility) can also control macroscopic phase behavior even when the driving noises remain mutually uncorrelated. In the context of wealth dynamics, this identifies the volatility configuration as an additional control dimension for inequality formation and condensation.
 
\begin{acknowledgments} 
This research was supported by the Basic Science Research Program through the National Research Foundation of Korea (NRF) (KR) [Grant No. NRF-RS-2026-25489888~(J.H. and M.H.) and Grant No.NRF-RS-2025-00514776~(J.H. and H.J.)].
\end{acknowledgments} 

\section*{DATA AVAILABILITY}
There are no publicly available research data or software supporting this manuscript. Requests for further information or data should be sent to the authors. 

\begin{appendix}

\section{\label{detail} DETAILS OF NUMERICAL SIMULATIONS}

We describe the numerical procedure used to estimate the effective tail exponent $\hat{\gamma}$. A linear fit to the double-logarithmic plot of $\rho(c)$ is highly sensitive to the histogram binning and does not provide estimates of $\hat{\gamma}$ consistent with the control parameter $\Lambda$. We use the maximum-likelihood estimator (MLE) for the Pareto distribution as follows:
\begin{align}
\hat{\gamma}=1+n\left[\sum_{i=1}^{n}\ln\left(\frac{c_i}{c_{\min}}\right)\right]^{-1},
\end{align}
where $c_i$ denotes the normalized wealth satisfying $c_i>c_{\min}$, $c_{\min}$ is the lower cutoff for the tail, and $n$ is the number of samples. To estimate $c_{\min}$, we use the Kolmogorov--Smirnov (KS) statistic~\cite{clauset2009power}. In summary, we determine $c_{\min}$ using a goodness-of-fit test based on the KS statistic and calculate $\hat{\gamma}$ using the MLE for the Pareto distribution for each run of the HBM simulation. In the main text, Fig.~\ref{fig3} shows $\hat{\gamma}$ averaged over 128 runs at $t=10^5$.

For network generation, we retain only the giant connected component (GCC), since disconnected nodes do not exchange wealth and would otherwise distort the stationary normalized wealth distribution. In an ER network, the degree distribution converges to a Poisson distribution as $N\to\infty$:
\[
p_{\mathrm{ER}}(k)=\frac{\lambda^k e^{-\lambda}}{k!},
\]
where $\lambda$ is the mean degree. The degree distribution of the GCC is
\[
p_{\mathrm{GCC}}(k)=\frac{\lambda^k e^{-\lambda}(1-u^k)}{k!(1-u)}.
\]
Here $u$ satisfies the self-consistent equation $u=e^{-\lambda(1-u)}$. The mean degree of the GCC is
\begin{align}
\langle{k}\rangle_{\mathrm{GCC}}=\lambda-W_0(-\lambda e^{-\lambda}),
\end{align}
where $W_0$ is the zeroth branch of the Lambert $W$ function, and $\langle{k}\rangle_{\rm GCC}$ is always larger than $\lambda$. Therefore, to obtain the target mean degree $\langle{k}\rangle_{\rm GCC}=4$ in numerical simulations, the ER network generation parameter must be chosen slightly below 4. Numerically, $\lambda\approx3.915$ gives $\langle{k}\rangle_{\mathrm{GCC}}=4$.

\begin{figure}[]
\includegraphics[width=\columnwidth]{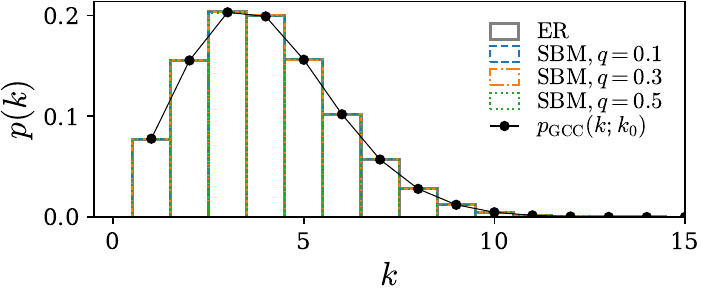}
    \caption{Degree distributions of giant connected components (GCCs) of ER networks and SBMs for various $q$. Black dots are the theoretical probability mass function for the GCC of the ER network with $k_0=3.915$. Colored histograms show the degree distribution obtained from 100 independent random network generations. For all cases, the mean degree is $\langle{k}\rangle=4.00\pm0.027$.}
    \label{fig-A1}
\end{figure}

The same preprocessing is applied to the SBM. After restricting networks to their GCCs, the ER and SBM networks used in numerical simulations have nearly identical mean degrees, degree distributions, and average path lengths. The corresponding SBMs yield almost the same wealth distributions and effective tail exponents $\hat{\gamma}$ in the homogeneous limit of $\Delta\beta\to 0$ [see Fig.~\ref{fig1}(b) in the main text]. Since the ER network and the SBM have almost identical degree distributions, their GCCs also have almost identical degree distributions regardless of $q$ (see Fig.~\ref{fig-A1}).
%
\begin{figure*}[]
\includegraphics[width=0.95\textwidth]{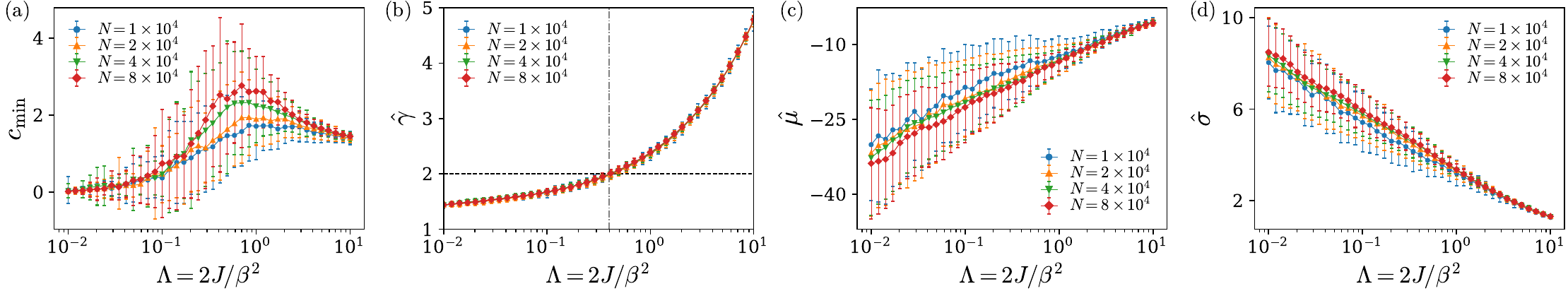}
    \caption{Estimated fitting parameters for tail distributions: (a),(b) the power-law parameters $c_{\min}$ and $\hat{\gamma}$, and (c), (d) the truncated log-normal parameters $(\hat{\mu}, \hat{\sigma})$. The cutoff $c_{\min}$ is determined by the KS statistic, while $\hat{\gamma}$ and $(\hat{\mu}, \hat{\sigma})$ are estimated by MLE. For the truncated log-normal fit, the same $c_{\min}$ is used as the lower truncation point. The normalized wealth distributions $\rho(c)$ are obtained from HBM simulations on ER networks with $\langle{k}\rangle =4.00\pm 0.027$. All dots and error bars represent the mean and standard deviation, respectively, over 128 independent runs.}
    \label{fig-A2}
\end{figure*}

\section{\label{complementary} 
COMPLEMENTARY NUMERICAL SIMULATIONS}

\subsection{Power-law test and size dependence}

We examine the statistical evidence for power-law tails and their robustness across parameter regimes. Bouchaud and M{\'e}zard~\cite{bouchaud2000wealth} reported power-law-tailed wealth distributions in random networks with a mean degree $\langle{k}\rangle=4$. Garlaschelli~\cite{garlaschelli2004wealth} discussed the relevance of log-normal behavior in sparsely connected clusters, whereas Souma {\it et al.}~\cite{souma2001small} showed that shortcuts in small-world networks can generate power-law tails even at a sparse and fixed mean degree. Motivated by these observations, we go beyond the visual inspection of double-logarithmic plots and compare the candidate tail distributions using a log-likelihood ratio test~\cite{vuong1989likelihood,clauset2009power}.

For the GCC of an ER network with a mean degree $\langle{k}\rangle=4$, whether the tail is better described by a power-law or a truncated log-normal distribution depends on the control parameter $\Lambda$ and the system size $N$. For $N=10^4$, the power-law fit is preferred for $\Lambda\geq1$, while for $N=8\times10^4$, it is already preferred for $\Lambda\geq0.588$. Thus, the parameter region favoring the power-law fit broadens as $N$ increases. The same trend is observed in the HBM model on the SBM. For smaller $\Lambda$, the finite-$N$ tail is better fitted by a truncated log-normal distribution. However, the corresponding MLE estimates $(\hat{\mu},\hat{\sigma})$ clearly depend on the system size, whereas the effective tail exponents $\hat{\gamma}$ collapse well and show only weak dependence on the system size (see Fig.~\ref{fig-A2}).

These results suggest that in the small-$\Lambda$ regime, the truncated log-normal fit may reflect finite-size rounding rather than the asymptotic tail form. As $N$ increases, the tail extends to large $c$, and the fitted log-normal tail becomes progressively flatter, while $\hat{\gamma}$ remains approximately stable. This behavior is consistent with a crossover to a power-law scaling in the thermodynamic limit $N\to\infty$. We therefore interpret $\hat{\gamma}$ as an effective scaling exponent, without claiming that finite-$N$ samples follow a pure power-law over the entire fitted range. Under this interpretation, the cases with $\hat{\gamma}\leq2$ are consistent with condensation in the thermodynamic limit.

\subsection{Wealth distribution and asymmetric wealth shares in the HBM model}

\begin{figure}[b]
\includegraphics[width=\columnwidth]{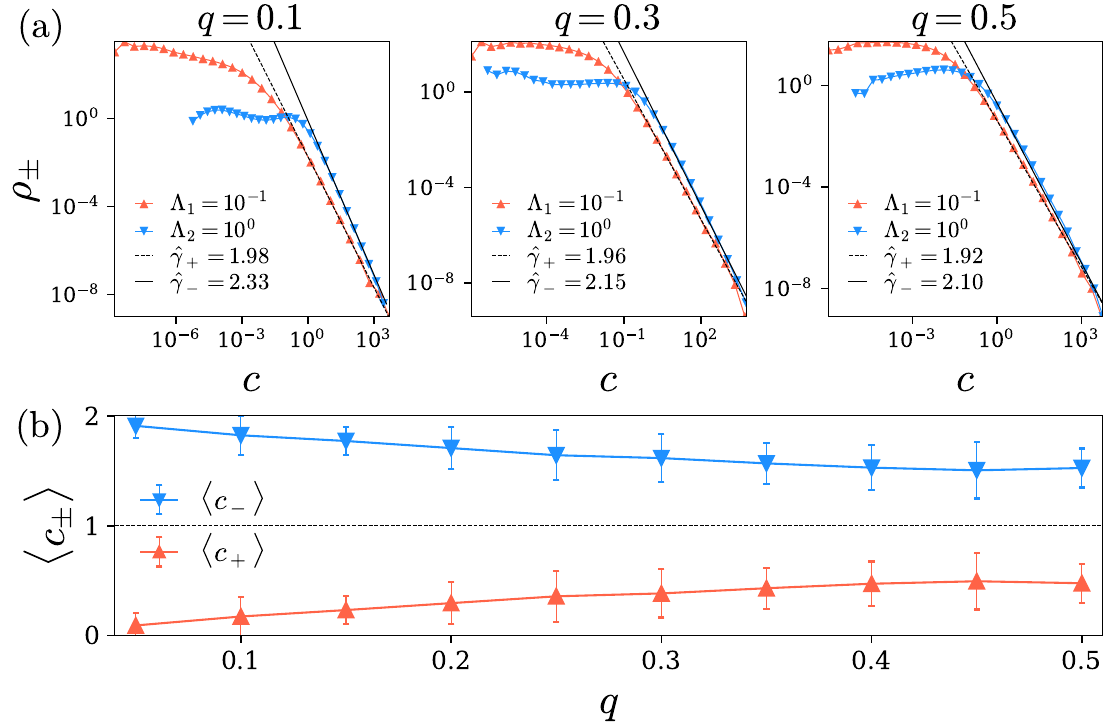}
    \caption{(a) Double-logarithmic scaled plots of group-wise normalized wealth distributions $\rho_{\pm}(c)$ in the HBM model on the SBM for various $q=\{0.1, 0.3, 0.5\}$. (b) Corresponding group-wise wealth shares $\langle c_{\pm} \rangle$ against $q$. For all cases, $(\Lambda_1,\Lambda_2)=(0.1,1)$. All results are averaged over 128 independent runs, and error bars indicate standard deviations.}
    \label{fig-A3}
\end{figure}

Figure~\ref{fig-A3} shows the normalized wealth distributions and group-wise wealth shares of two volatility groups. The comparison of group-wise distributions indicates that both the effective tail exponent of the aggregate distribution and the wealth shares are primarily governed by the low-volatility group. As $q$ increases, the difference between the two groups in the tail sector decreases, and the disparity in their wealth shares is reduced accordingly.

When the effective tail exponent satisfies $\hat{\gamma}\ll 2$, the system lies deep in the condensation regime, where even a small difference in the tail populations is strongly amplified in the wealth shares. Consequently, the reduction in the wealth-share disparity with increasing $q$ remains negligible in this regime. By contrast, as $\hat{\gamma}$ increases, this condensation-induced amplification weakens, and the wealth-share disparity decreases more noticeably with increasing $q$.

\subsection{\label{Appendix-C3} Isolating the effect of volatility--volatility assortativity on a fixed network}

We primarily considered a symmetric two-block SBM to easily control assortativity $\mathcal{A}$. The SBM is equivalent to an ER network at $q=1/2$, but not at other values of $q$. Consequently, numerical simulations in the main text use different network realizations for each $\mathcal{A}$. To isolate the effect of $\mathcal{A}$ while keeping the network fixed, we employ a random pair-swapping algorithm. Such correlation control procedures are necessary for investigating the effect of metadata configurations in empirical networks.

The procedure consists of the following three steps:
(i) Randomly select two nodes and swap their assigned volatility values. (ii) Accept the swap if it changes $\mathcal{A}$ in the desired direction, and otherwise reject it. (iii) Repeat these steps until the target value of $\mathcal{A}$ is reached. This procedure generates configurations with different values of $\mathcal{A}$ on the same network. However, changing $\mathcal{A}$ may also alter other correlations. 

A relevant example is the degree--volatility correlation ($DV$), defined as
\begin{align}
DV=\frac{{\rm Cov}(k,\beta)}{\sqrt{{\rm Var}(k){\rm Var}(\beta)}},
\label{eq-C1}
\end{align}
where $k$ and $\beta$ denote the degree and volatility of a node, respectively. In the symmetric two-block SBM, interblock connections are random, so the volatility assignment is not biased by degree and $DV\approx0$, regardless of $q$. By contrast, random pair swapping can change $DV$ and $\mathcal{A}$ simultaneously. Therefore, we use a conditioned random pair-swapping algorithm that accepts a swap only when the resulting configuration satisfies $|DV|<\epsilon$. This allows us to vary $\mathcal{A}$ while keeping $DV$ approximately fixed at zero.

\begin{figure}[b]
\includegraphics[width=\columnwidth]{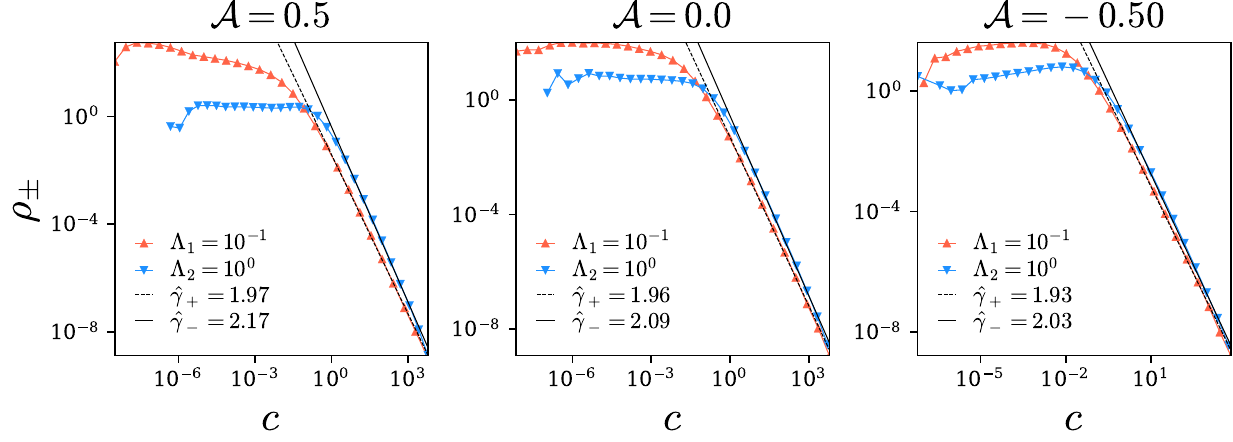}
    \caption{Double-logarithmic scaled plots of group-wise normalized wealth distribution $\rho_{\pm}(c)$ in the HBM model on the ER network with a mean degree $\langle{k}\rangle=4.00$ for various $\mathcal{A}=\{0.5, 0.0, -0.5\}$. For each case, the volatility configuration is obtained by the conditioned random pair-swapping algorithm and the degree--volatility correlation is controlled as $|DV|<10^{-2}$. All results are obtained from 1280 ensembles.}
    \label{fig-A4}
\end{figure}

We apply the conditioned random pair-swapping algorithm to an ER network with $\langle{k}\rangle=4.00\pm0.027$, which imposes $|DV|<\epsilon=10^{-2}$. For a fixed network, the attainable range of $\mathcal{A}$ is structurally bounded~\cite{cinelli2020network}. After $100N$ swap attempts, the attainable minimum and maximum values are approximately $-0.5$ and $0.5$, respectively. Figure~\ref{fig-A4} shows the results of the HBM simulation for $\mathcal{A}\in\{0.5,0.0,-0.5\}$. These results are consistent with those in the main text, showing both the neutralization of the tail exponents and the dominance of the low-volatility group in the tail population.

\end{appendix}

\bibliography{ref-proof}

\begin{thebibliography}{18}
\bibitem [{SDE()}]{SDE-normalizedwealth}%
  \BibitemOpen
  \href@noop {} {}\bibinfo {note} {For the original SDE of BM model: $dC_i=\alpha C_idt+\beta C_idW_{t,i}+\sum_{j(\neq i)}\left(J_{ij}C_j-J_{ji}C_i\right)dt$, the normalized wealth is defined as $c_i=C_i/\bar{C}$ which always gives $\langle{c}\rangle=1$.}\BibitemShut {Stop}%
\bibitem [{deg()}]{degree-degree}%
  \BibitemOpen
  \href@noop {} {}\bibinfo {note} {We also note that volatility-volatility correlation $\mathcal{A}$ differs from the degree--degree correlations~\cite{newman2002assortative,yang2017lower}}\BibitemShut {NoStop}%
\end{thebibliography}

\end{document}